\documentclass[
aps,
pra,
twocolumn,
superscriptaddress,
floatfix
longbibliography
]{revtex4-1}
\usepackage[dvipdfmx]{graphicx}
\usepackage{verbatim}
\usepackage{epsf}
\usepackage{natbib}
\usepackage{dcolumn}
\usepackage{bm}
\usepackage{color} 
\usepackage{ulem}
\usepackage{amsmath}
\usepackage{amssymb}
\usepackage[
colorlinks=true,
citecolor=blue,
urlcolor=magenta,
setpagesize=false]{hyperref}
\usepackage{textcomp}

\newcommand{\dagga}{{\phantom{\dagger}}}

\newcommand{\phihfUtwo}{0.0098(3) }
\newcommand{\phihfUthree}{0.0222(6) }
\newcommand{\phihfUfour}{0.0368(2) }

\newcommand{\phiqfUtwo}{0.0009(3) }
\newcommand{\phiqfUthree}{0.0088(1) }
\newcommand{\phiqfUfour}{0.0266(2) }

\newcommand{\MUtwo}{0.122(1) }
\newcommand{\MUthree}{0.183(2) }
\newcommand{\MUfour}{0.2347(4) }

\newcommand{\HFtwo}{0.1881 }
\newcommand{\HFthree}{0.2830 }
\newcommand{\HFfour}{0.3453 }

\newcommand{\MxyUtwo}{0.030(4) }
\newcommand{\MxyUthree}{0.094(1) }
\newcommand{\MxyUfour}{0.163(1) }

\newcommand{\phihfUtwovmc}{0.01994(4) }
\newcommand{\phihfUthreevmc}{0.0497(4) }
\newcommand{\phihfUfourvmc}{0.0758(5) }

\newcommand{\phiqfUtwovmc}{0.00095(25) }
\newcommand{\phiqfUthreevmc}{0.0143(1) }
\newcommand{\phiqfUfourvmc}{0.0378(4) }

\newcommand{\MUtwovmc}{0.1412(1) }
\newcommand{\MUthreevmc}{0.2230(6) }
\newcommand{\MUfourvmc}{0.2752(6) }

\newcommand{\MxyUtwovmc}{0.031(4) }
\newcommand{\MxyUthreevmc}{0.1196(6) }
\newcommand{\MxyUfourvmc}{0.194(1) }

\newcommand{\Enhftwo}{-2.1755(3) }
\newcommand{\Enhfthree}{-2.5014(3) }
\newcommand{\Enhffour}{-2.86016(9) }

\newcommand{\Enhftwovmc}{-2.16848(1) }
\newcommand{\Enhfthreevmc}{-2.49007(1) }
\newcommand{\Enhffourvmc}{-2.84769(5)}

\newcommand{\Enqftwo}{-1.4652(5) }
\newcommand{\Enqfthree}{-1.5669(5) }
\newcommand{\Enqffour}{-1.6920(2) }

\newcommand{\Enqftwovmc}{-1.4615(2)}
\newcommand{\Enqfthreevmc}{-1.55820(6) }
\newcommand{\Enqffourvmc}{-1.68016(4)}

\begin{document}

\title{
Study of the superconducting order parameter in the negative-$U$ 2D-Hubbard model by grand-canonical  twist-averaged boundary conditions 
}

\author{Seher~Karakuzu}
\affiliation{International School for Advanced Studies (SISSA), Via Bonomea 265, 34136, Trieste, Italy}

\author{Kazuhiro~Seki}
\affiliation{International School for Advanced Studies (SISSA), Via Bonomea 265, 34136, Trieste, Italy}
\affiliation{Computational Materials Science Research Team, RIKEN Center for Computational Science (R-CCS),  Hyogo 650-0047,  Japan}
\affiliation{Computational Condensed Matter Physics Laboratory, RIKEN Cluster for Pioneering Research (CPR), Saitama 351-0198, Japan}

\author{Sandro~Sorella}
\affiliation{International School for Advanced Studies (SISSA), Via Bonomea 265, 34136, Trieste, Italy}

\begin{abstract}
  By using variational Monte Carlo and auxiliary-field quantum Monte Carlo methods, 
  we perform an accurate finite-size scaling  of the $s$-wave superconducting order parameter and 
  the pairing correlations for the negative-$U$  Hubbard model at zero temperature in the square lattice. 
  We show that the twist-averaged boundary conditions (TABCs) are extremely important to 
  control  finite-size effects and to achieve smooth and accurate extrapolations to the thermodynamic limit.  
  We also show that TABCs is much more efficient in the grand-canonical ensemble 
  rather than  in the standard canonical ensemble with fixed number of electrons.  
  The superconducting order parameter as a function of the doping is presented for several values of $|U|/t$ and 
  is found to be significantly smaller than the mean-field BCS estimate already for moderate couplings. This reduction is understood 
  by  a variational ansatz able to describe the low-energy behaviour 
  of the superconducting 
  phase, by means of a suitably chosen Jastrow factor including  long-range density-density 
  correlations.
\end{abstract}

\date{\today}

\maketitle

\section{Introduction}
In recent years the numerical simulations has achieved a constantly  increasing 
impact in  theoretical and experimental condensed matter physics, because, 
on one hand it allows reliable  solutions of correlated models which cannot be solved analytically~\cite{PRX_Hub} and, on the other hand, the quest 
of accurate benchmark results is now becoming of fundamental importance.
Indeed the emergent collective properties of quantum many-body systems, such as Bose-Einstein condensation (BEC) and superconductivity, can be now 
probed directly  by 
  clean  and realistic  representations of Hubbard-like 
 lattice models with 
 ultracold atoms trapped in optical lattices~\cite{Inouye1998,Courteille1998,Greiner2003,Bloch2008}. 

The negative $U$ two-dimensional (2D) Hubbard model is a very simple  model 
of electrons subject to an attractive interaction on a lattice.
It is clearly relevant for 
  studying the standard mechanism of superconductivity within the 
Bardeen-Cooper-Schrieffer  (BCS) theory~\cite{BCS1957,Leggett1980,AHMmicnas,Nozieres1985}. 
At finite temperatures, the phase diagram of the model 
has been investigated by quantum Monte Carlo (QMC)
~\cite{Scalettar1989negU,AHMScalapino1991,Singer1996,Singer1998,AHMScalapino1992} 
as well as by dynamical-mean-field theory calculations~\cite{Capo2005,Capo2005-2}. 
Normal (non-superconducting) state properties have been studied 
via finite-temperature Monte Carlo calculations~\cite{MRanderia2001}, by focusing mainly on the BCS-BEC crossover, and recently~\cite{Otsuka2018}
the zero temperature quantum critical 
point between a metal and a superconductor was also satisfactorily described, thanks to large scale simulations,
 nowadays possible with  modern supercomputers and the excellent algorithmic  performances of QMC.
At zero temperature, 
ground-state properties of the model have been also studied  
by variational Monte Carlo (VMC) calculations 
as a function of the  interaction strength for several electron fillings~\cite{Yokoyama,Yokoyama2002}. 

As well known, the main purpose in the numerical simulation is 
to reach a controlled and accurate  thermodynamic limit of a model system 
with a sequence of calculations with increasing number of electrons.
This task may be particularly difficult especially in the weak-coupling 
($|U|/t \ll 8$) regime,  because in this limit the location of the 
Fermi surface plays a crucial 
role.  Indeed, the results obtained with conventional periodic-boundary conditions 
(PBC) may significantly depend on the particular location 
of the allowed finite-size momenta, resulting in very difficult, if not impossible,
extrapolations to the thermodynamic limit. 
The  drawback of PBC is well known and  represents an important 
limitation of most numerical techniques dealing with fermions.
Indeed, an early projector Monte Carlo study has shown 
strong finite-size effects on superconducting pairing correlations~\cite{Bormann1991}. 
Obviously, when the Fermi surface is particularly simple 
such as the 2D-half-filled Hubbard model with its perfectly nested Fermi surface, this problem 
is less severe but, at weak coupling, even this particular simple case may deserve some attention.

In order to control the finite size effects discussed above, twist-averaged boundary conditions (TABCs) 
have been introduced for Monte Carlo simulations
on  lattice model~\cite{gros_exact_diag,gros_gtabc,Koretsune2007,Karakuzu2017,Me2016,SZhang2016}  
and in continuum systems~\cite{CLin,Holzmann2006}. 
Within TABCs, physical quantities are estimated by 
averaging them  over several twisted-boundary conditions~\cite{Poilblanc1991}, 
rather than limiting the calculation to a single twist, such as PBC. 
In this way,  
TABCs can substantially reduce finite-size effects~\cite{CLin,Koretsune2007,Karakuzu2017,Me2016,SZhang2016}, at the expense of performing several independent 
calculations with several twists. 
In QMC this overhead is not even relevant because, at given computational resources,  the statistical errors of the twisted-averaged quantities do not grow 
with the number $N_{\rm TABC}$ of twists.
Thus this method is particularly appealing within QMC and, quite recently, 
is becoming widely used for the study of strongly correlated systems. 
On the other hand, in 
a recent work~\cite{Fakher2017}, by using
finite-temperature determinant quantum Monte Carlo without TABCs, 
the convergence of physical quantities to the thermodynamic limit 
have been examined 
for the canonical ensemble (CE) and the grand-canonical ensemble (GCE). 
It has been shown that GCE provides a convergence faster  than CE. 
There are several reasons why this should happen. The simplest one 
is that only by allowing 
 the fluctuations of the particle number  one  can ensure that the $U=0$ Gibbs-free energy 
 coincides with the one in the thermodynamic limit~\cite{gros_exact_diag}. 
On the other hand, at zero temperature this technique is equivalent to occupy only the electronic 
states within the given Fermi surface, and this may explain why 
it is so important for fermionic systems, at least in the weakly correlated regime.

Since the size effects are certainly more  pronounced at zero temperature 
and weak coupling, it is important  to 
explore and benchmark systematically  more efficient ways to reduce the finite-size error, in order 
to assess with some confidence 
the behavior  of the superconducting  order parameter -- non zero in 2D only at zero temperature -- 
in the BCS regime.

In this paper, we examine finite-size effects on the $s$-wave superconducting 
order parameter and pairing  correlations in the 2D negative-$U$ Hubbard model 
by using VMC and AFQMC methods at zero temperature. 
We introduce a combination of TABCs with GCE sampling technique at zero temperature 
and show that the finite-size effects are more efficiently reduced in GCE than in CE.

The rest of this paper is organized as follows. 
In Sec.~\ref{sec2}, we describe the negative-$U$ Hubbard model, VMC and AFQMC methods, 
and TABCs on a 2D square lattice.
In Sec.~\ref{sec3}, we present numerical results of the $s$-wave order parameter and the 
pairing correlation functions for the entire doping range at several values of the 
interaction strength. 
In Sec.~\ref{sec4}, we draw our conclusions   and   discuss the implications of the present method 
for future works.

\section{Model and Method} \label{sec2}
\subsection{Negative-$U$ Hubbard model}
The Hamiltonian of the negative-$U$ Hubbard model is given as~\cite{Hubbard238}
\begin{eqnarray}
  {\cal H} = {\cal H}_{\cal K} + {\cal H}_{\cal V}
  \label{eq:Hmodel}
\end{eqnarray}
with 
\begin{eqnarray}
  &&{\cal H}_{\cal K} = -t \sum_{\langle i,j \rangle , \sigma} 
  \left(c^\dag_{i\sigma} c^\dagga_{j\sigma}   + {\rm H. c.} \right) 
  - \mu \sum_{i\sigma} n_{i\sigma}, \\
  &&{\cal H}_{\cal V} =  U \sum_{i} n_{i\uparrow} n_{i\downarrow},
\end{eqnarray}
where $t$ is the hopping integral and  $\langle i,j \rangle$ indicate
nearest-neighbors
on a square lattice with $N$ sites, $c^\dag_{i\sigma}$ ($c_{i\sigma}$) creates (destroys) an electron 
with spin $\sigma( = \uparrow, \downarrow)$ on the site $i$, and $n_{i \sigma} = c^\dag_{i\sigma} c^\dagga_{i\sigma} $. 
$U<0$ is the Hubbard interaction term which, in this paper, is considered to be negative and $\mu$ is the chemical potential. 
Hereafter we set $t$ and the lattice constant, both equal to one.

\subsection{Variational Monte Carlo}
In order to study the negative-$U$ Hubbard Model defined in Eq.~(\ref{eq:Hmodel}), 
we employ the VMC method. 
As a variational many-body wavefunction for VMC, 
we use a Jastrow-Slater wavefunction of the form 
\begin{equation}\label{eq:wf}
  |\Psi\rangle = {\cal J} |\Psi_{\rm T} \rangle, 
\end{equation}
where $\cal J$ is the density-density Jastrow correlator defined by 
\begin{equation}
  {\cal J}= \exp \left ( -\frac{1}{2} \sum_{i,j} v_{i,j} n_{i} n_{j} \right ), \label{eq:Jas} 
\end{equation}
with $n_{i}=\sum_{\sigma} n_{i\sigma}$ and 
$v_{i,j}$ being the variational parameters 
which are assumed to depend only on the distance between the sites $i$ and $j$. 
It is particularly important to consider in this study a Jastrow factor 
where the pseudopotential $v_{i,j}$ is non zero even when the two lattice points are at very large distance $d$, 
because  in a superconductor the pseudopotential should decay as $ \simeq 1/d$~\cite{capello}, in order to define a physical 
wavefunction with correct charge fluctuations at small momenta.
Moreover, when the fluctuations of the number of particles is considered, 
a fugacity term $\exp(-f \sum\limits_i n_i)$ has to be added to Eq.~(\ref{eq:Jas}). 
At half filling the fugacity  is 
determined by the condition that Eq.~(\ref{eq:Jas}) remains unchanged (up to a constant) for the 
particle-hole symmetry:
\begin{equation}
c_{i \sigma} \to (-1)^{x_i+y_i} c^{\dag}_{i -\sigma}
\label{eq:ph}
\end{equation}
where $x_i,y_i$ are the lattice  coordinates of the site $i$. 
This implies that $f= { 1 \over N} \sum_{i,j} v_{i,j}$ after a straightforward calculation.

The antisymmetric part of the wavefunction, $|\Psi_{\rm T} \rangle$, 
is obtained from the ground state of a mean-field (MF)  Hamiltonian ${\cal H}_{\rm MF}$ that 
contains the electron hopping, chemical potential and singlet $s$-wave pairing terms;
\begin{eqnarray}
  {\cal H}_{\rm MF} &=& -t \sum_{\langle i,j \rangle,\sigma} \left( c^\dag_{i\sigma} c^\dagga_{j\sigma} + {\rm H. c.}\right)  
  - \mu_{\rm BCS} \sum_{i\sigma} n_{i\sigma}  \notag \\
  &+&\Delta_{\rm 0} \sum_{i} \left( c^\dag_{i \uparrow} c^\dag_{i \downarrow} + {\rm H. c.} \right),  
  \label{eq:MFEqn}
\end{eqnarray}
where $\mu_{\rm BCS}$, and  $\Delta_{0}$ are variational parameters. 
All the variational parameters $v_{i,j}$, $\mu_{\rm BCS}$, and  $\Delta_{0}$ are optimized 
via stochastic-reconfiguration technique by minimizing the variational expectation value of the  energy~\cite{Sorella2005}. 

In order to do the Monte Carlo integration, configurations, where electrons have 
a definite position and spin quantization axis $S^z_i=\pm 1/2$,  are sampled through 
Markov chains and proposed moves are accepted or rejected with the Metropolis algorithm. 
In particular it is possible  to consider  the moves (hoppings) 
defined by the Hamiltonian of the system of interest. With this limitation the VMC  conserves the total number of 
particles and the total projection $S^z_{\rm tot}=\sum\limits_i S^z_i=0$ 
of the spin in the chosen quantization 
axis.  Thus, these kind of projections 
are implicitly assumed in Eq.~(\ref{eq:wf}).
In this work we have considered also moves that change the number of particles 
(remaining in the $S^z_{\rm tot}=0$ subspace). With this in mind, one can extend the sampling from  CE to  GCE by enlarging the Hilbert space, where the former consists of local moves conserving the particle number while the latter includes moves allowing fluctuations of the particle number.

\subsection{Auxiliary-field quantum Monte Carlo}
In order to test the relevance of the correlated Ansatz in Eq.~(\ref{eq:Hmodel}) for VMC, 
we also employ the AFQMC method. 
AFQMC is based on the idea that 
the imaginary-time propagation of a trial wavefunction $|\Psi_{\rm T} \rangle$ 
with a long-enough projection time can project out 
the exact ground-state wavefunction $|\Psi_0 \rangle$, 
provided that the trial wavefunction is not orthogonal to 
the exact ground-state wavefunction, i.e., 
$\langle \Psi_{\rm T} | \Psi_{0} \rangle \not = 0$~\cite{Sorella89}. 
AFQMC suffers from the negative-sign problem for $U>0$ 
if the particle-hole symmetry is broken.  
However, for the case of the negative-$U$ Hubbard model, 
there is no sign problem whenever the number of 
up-spin particles equals the one of  down-spin particles~\cite{becca_sorella_2017}.

We define a pseudo-partition function by~\cite{becca_sorella_2017} 
\begin{equation}
  \mathcal{Z} = \langle \Psi_{\rm T}|e^{-\beta \cal H}|\Psi_{\rm T} \rangle 
  = \langle\Psi_{\rm T}|(e^{-\Delta \tau \cal H})^{2T} |\Psi_{\rm T} \rangle, 
\end{equation}
where $\beta$ is the projection time and is discretized into $2T$ time slices, 
i.e. $\Delta \tau ={ \beta \over 2 T} $  
in the RHS of the above equation. 
Then the ground-state expectation value of an operator $\cal O$ can be written as 
\begin{equation}
  \frac{\langle \Psi_{0} | {\cal O}| \Psi_{0} \rangle }{\langle \Psi_{0}|\Psi_{0} \rangle} =
  \lim_{T\to\infty}  \frac{\langle \Psi_{\rm T} |(e^{-\Delta \tau \cal H})^{T} {\cal O} (e^{-\Delta \tau \cal H})^{T}|\Psi_{\rm T} \rangle }{\mathcal{Z}}.
\end{equation}

Since the interaction part of the Hamiltonian $\cal H_{V}$ consists of 
a two-body term and does not commute with the kinetic part $\cal H_{K}$, 
the imaginary-time propagator $e^{-\Delta \tau {\cal H}}$ requires the following  manipulation.  
First, in order to factorise the Hamiltonian into the 
interaction and kinetic parts in the exponential, 
we use the symmetric Trotter-Suzuki decomposition~\cite{Trotter1959,Suzuki1976}
\begin{equation}
  e^{-\Delta \tau \cal H} = 
  e^{-\frac{\Delta \tau}{2} {\cal H}_{\cal K}} 
  e^{-\Delta \tau {\cal H}_{\cal V} } 
  e^{-\frac{\Delta \tau}{2} {\cal H}_{\cal K}} + O(\Delta \tau^{3}), 
\end{equation}
where $O(\Delta \tau^{3})$ is the systematic error due to the time discretization. 
Since there are $2T$ number of slices, 
the errors are accumulated and the resulting systematic error is $O(\Delta \tau^{2})$. 
We set the projection time to be $\beta= 3L$ with a fixed imaginary-time discretization $\Delta \tau= 0.1$. 
It has been shown that $\Delta \tau=0.1$ is small enough to accurately determine the ground-state 
phase diagram of the honeycomb-lattice Hubbard model in the weak-coupling regime~\cite{Sorella2012}.  
Then, we write the interaction term as a superposition of one-body propagators  by means of the well established  Hubbard-Stratonovich transformation~\cite{Hubbard1959,Stratonovich1957}. 
Hirsch pointed out that since the occupation numbers are only 0 or 1 for fermions, 
one can introduce Ising-like discrete fields, $s_{i} = \pm 1$ \cite{Hirsch1985}, such that
\begin{eqnarray}
&&\prod_{i} e^{\Delta\tau |U| n_{i,\uparrow}n_{i,\downarrow}} \notag \\
&=& \prod_{i} \frac{1}{2} e^{\frac{\Delta\tau |U|}{2}(n_{i\uparrow} + n_{i\downarrow} - 1 )} 
\sum_{s_{i} =  \pm 1}   e^{s_{i}\gamma(n_{i\uparrow} + n_{i\downarrow} - 1)}, 
\end{eqnarray}
where $\cosh{\gamma}=e^{\frac{\Delta\tau |U|}{2}}$. 
The summation over the auxiliary fields $\{s_{i}\}$ is performed by the Monte Carlo sampling. 
For AFQMC, the sampling is done via Markov chains based on  local field-flip sequential updates. 

When $\Delta_{0}=0$, 
the trial wavefunction $|\Psi_{\rm T} \rangle$ is constructed by 
filling the lowest-lying orbitals for a fixed particle number and therefore the sampling is done in CE. 
When $\Delta_{0}\not=0$, the sampling is automatically done in GCE, 
and the desired particle number is obtained by tuning the chemical potential $\mu$. 
To determine the chemical potential for a desired particle number, we use the Newton-Raphson method,  
where the derivative of the particle number with respect to the chemical potential 
($\propto$ fluctuation of the particle number) 
is calculated directly by AFQMC simulations.  
In AFQMC, when using a single twist (and no TABCs), the trial wavefunction $|\Psi_{\rm T} \rangle$ 
 is the free electron ground state of ${\cal H}_{\cal K}$, satisfying the closed-shell condition 
in order to preserve all the  symmetries of the Hamiltonian. 
On the other hand, for the GCE calculations at finite doping we have used trial 
wavefunctions obtained by VMC optimization of the bare chemical potential  $\mu_{\rm BCS}$  and 
a small $s$-wave pairing [$\sim O(10^{-2}t)$], which allows particle fluctuations within CGE.

\subsection{Twist-averaged boundary conditions}
In the case of weakly correlated systems, 
size effects are most pronounced and calculations of observables 
with a single boundary condition such as 
PBC or anti-periodic-boundary condition (APBC) 
may have serious difficulties in determining the correct thermodynamic limit. 
In order to mimic the Brillouin zone of 
the thermodynamic limit, TABCs have been proposed and indeed it has been shown 
that TABCs eliminate one-body error very successfully~\cite{Karakuzu2017,CLin, Me2016}. 

On a lattice, by explicitly indicating the coordinates of the site 
in the creation operators, i.e.,
$c^\dag_{i\sigma} \to c^\dag_{{\bf R}_i\sigma}$, where ${\bf R}_i=(x_i,y_i)$ denotes 
the coordinates of the site $i$ in the lattice, 
twisted-boundary conditions correspond to impose~\cite{Poilblanc1991}: 
\begin{equation}
  \begin{split}
    c^\dag_{{\bf R}_i+{\bf L_x}\sigma} = e^{i\theta^{\sigma}_{x}} c^\dag_{{\bf R}_i\sigma}, \\
    c^\dag_{{\bf R}_i+{\bf L_y}\sigma} = e^{i\theta^{\sigma}_{y}} c^\dag_{{\bf R}_i\sigma},  \label{eq:twistang}
  \end{split}
\end{equation}
where ${\bf L_x}=(L,0)$ and ${\bf L_y}=(0,L)$ are 
the vectors that define the periodicity of the cluster; 
$\theta^{\sigma}_{x}$ and $\theta^{\sigma}_{y}$ are 
two phases in the interval $(-\pi,\pi)$ determining 
the twists along $x$ and $y$ directions, respectively. 
The number of  sites is given by $N=L^2$. 
In order to preserve time-reversal invariance of the singlet pairs, 
we imposed that $\theta^{\uparrow}=-\theta^{\downarrow}$ in both directions.

The expectation value of the operator ${\cal O}$ 
in TABCs is defined by 
\begin{equation}\label{eq:kaverage}
  \langle {\cal O} \rangle = \frac{1}{N_{\rm TABC}} \sum_{\theta} 
  \frac{\langle \Psi_{\theta}| {\cal O}_{\theta} |\Psi_{\theta}\rangle}
       {\langle \Psi_{\theta}|\Psi_{\theta}\rangle}.
\end{equation}
where $\theta=(\theta_x^\sigma,\theta_y^\sigma)$, 
${\cal O}_{\theta}$ is the operator corresponding to ${\cal O}$ 
under the boundary condition~Eq.~(\ref{eq:twistang}),  
$N_{\rm TABC}$ is the number of twist angles in the whole Brillouin zone, and 
$|\Psi_{\theta}\rangle$ is the wavefunction $|\Psi\rangle$ for VMC 
or $|\Psi_0\rangle$ for AFQMC, constructed by imposing 
the twisted-boundary conditions defined  in Eq.~(\ref{eq:twistang})  
to the trial wavefunction $|\Psi_{\rm T} \rangle$ as well as 
to the one-body part of the Hamiltonian. 
Note however that all the wavefunctions with different $\theta$ 
share the same variational parameters. In order to perform TABCs, 
we typically take $N_{\rm TABC}=1088$ points in the Brillouin zone.

\section{Results}\label{sec3}

\subsection{Size effects in mean-field approximation}

\begin{figure}[t!]
  \includegraphics[width=0.4\textwidth]{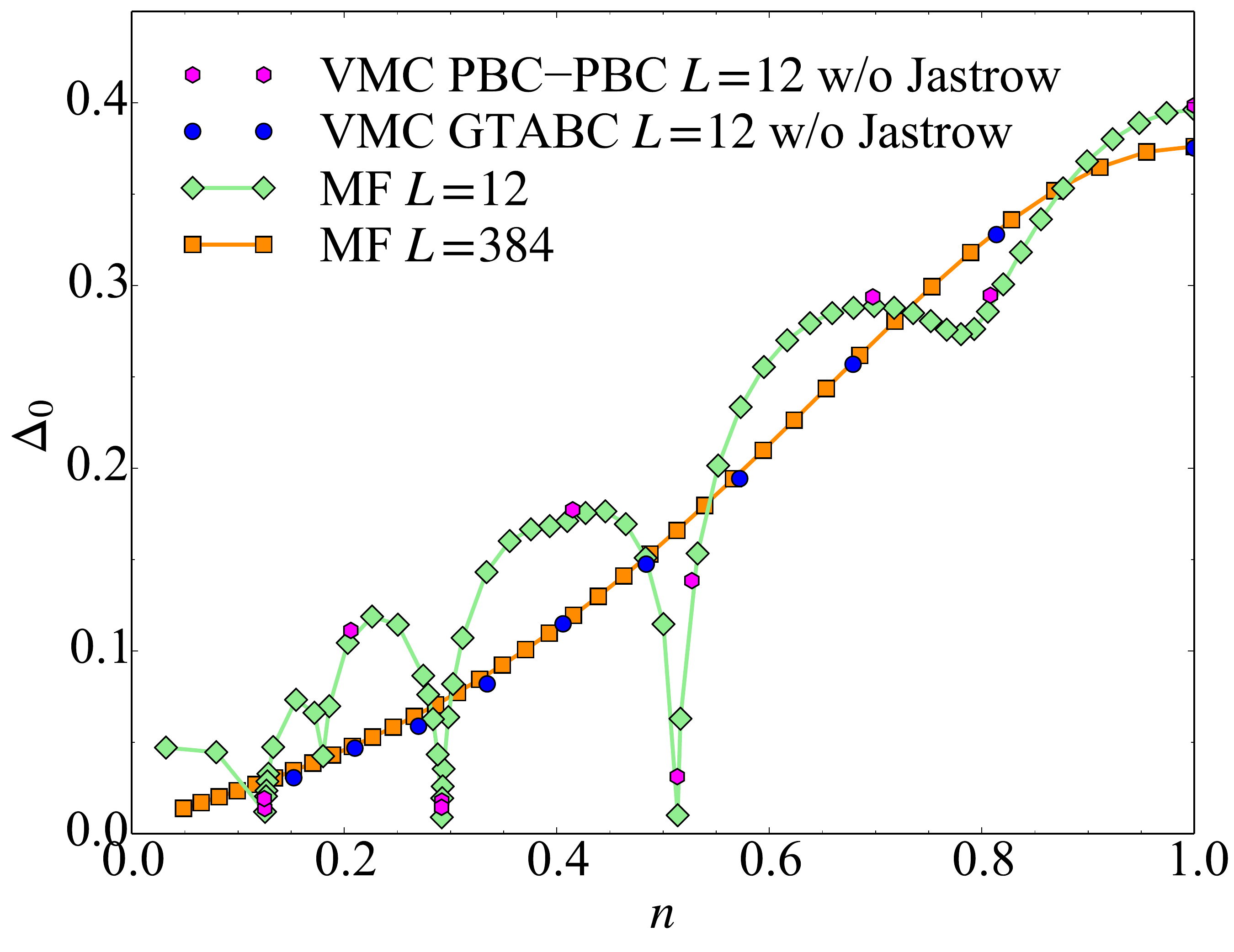}
  \caption{\label{fig:mf} 
    $s$-wave variational parameter $\Delta_0$ as a function of $n$ at $U=-2$. 
    Mean-field calculations are performed on $L=12$ and $L=384$ clusters,  
    and VMC calculations are done in GCE without Jastrow correlator on $L=12$ 
    with PBC and TABCs. 
    The error bars in the VMC results are smaller than the symbol sizes. 
 }
\end{figure}

Before investigating the finite-size effects in correlated systems, it is instructive to 
study the finite-size effects within the single-particle theory. 
For this purpose, we treat the negative-$U$ Hubbard model in Eq.~(\ref{eq:Hmodel})  
within the self-consistent mean-field approximation by 
decoupling the interaction term into the $s$-wave pairing terms. 

Figure~\ref{fig:mf} shows the $s$-wave superconducting order parameter $\Delta_0$ 
as a function of electron density $n$ ($n=1$ corresponds to the half filling) 
within the mean-field approximation at $U=-2$ for $L=12$ and $L=384$. 
We have confirmed that the order parameter does not depend on the system size for $L \geqslant 384$, 
implying that the results for $L=384$ can be considered very close to the thermodynamic limit.  
On the other hand, significant size effects, namely the oscillatory dependence on $n$, 
are observed for $L=12$. 

In order to test the accuracy of our VMC calculation, 
we have reproduced the above results  
by setting the Jastrow correlator $\mathcal{J}$ 
in Eq.~(\ref{eq:wf}) to be unity, i.e., $v_{i,j}=0$. 
The VMC calculations are performed for $L=12$, using  a single twist or   $32 \times 32$
twist angles in the whole Brillouin zone. Notice that, the latter case, 
corresponds, within a mean-field approach, to a single calculation with $L=384$ and PBC. 
The results obtained by VMC in GCE without Jastrow part are indeed in perfect agreement  with those obtained independently by the mean-field calculation.

\subsection{$s$-wave variational parameter}
\begin{figure}[t!]
  \includegraphics[width=0.4\textwidth]{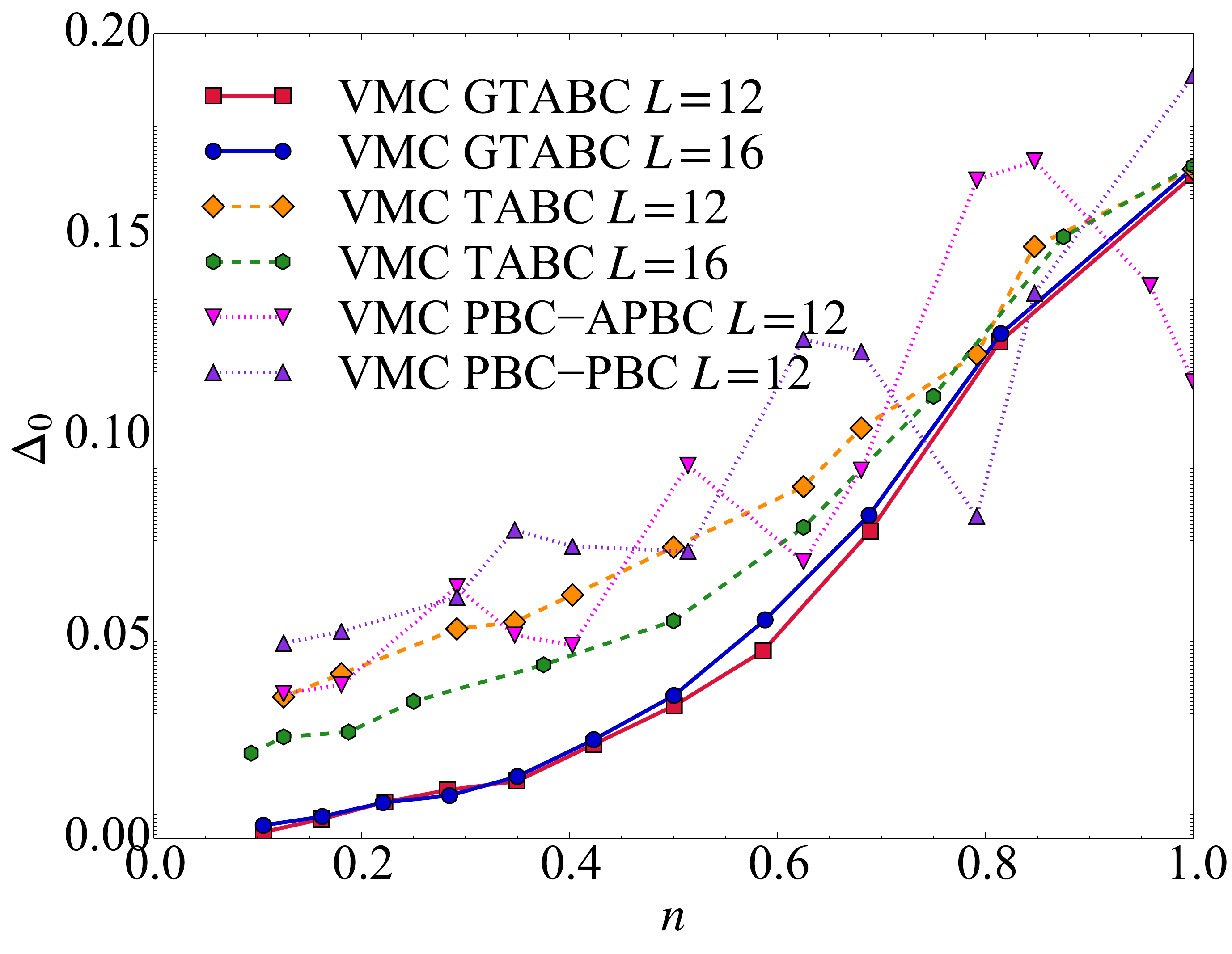}
  \caption{\label{fig:jastrow} 
    $s$-wave variational parameter $\Delta_0$ as a function of $n$ at $U=-2$ calculated by VMC. 
    The system size and boundary conditions used are indicated in the figure. 
    Here, 
    GTABC represents the grand-canonical twist-averaged boundary conditions, 
    TABC the canonical twist-averaged boundary conditions, 
    PBC-APBC stands for  PBC in one direction and APBC in the other one,
    whereas 
    PBC-PBC indicates PBC in both directions. 
    The error bars are smaller than the symbol sizes.
  }
\end{figure}

The mean-field results do not take into account the correlations between the electrons. 
The accuracy for treating the electron correlations can be improved 
by including the Jastrow factor in Eq.~(\ref{eq:wf}). 
Figure~\ref{fig:jastrow} shows the superconducting variational parameter $\Delta_0$ 
as a function of electron density $n$ within VMC for $L=12$ and $L=16$ with 
different boundary conditions and different ensembles. 
For a fixed system size of $L=12$, the results with a single boundary condition 
show oscillatory dependencies on $n$, similarly to the ones obtained within 
the mean-field approximation for $L=12$. 
With TABCs in both ensembles, the oscillatory dependencies are significantly reduced. 
By further increasing the system size to $L=16$, a sizable decrease 
of $\Delta_0$ is observed in CE especially for the low-density regime, 
while the change in GCE is almost negligible, indicating that the GCE shows much smaller 
size effects.

\begin{figure}[t!]
  \includegraphics[width=0.4\textwidth]{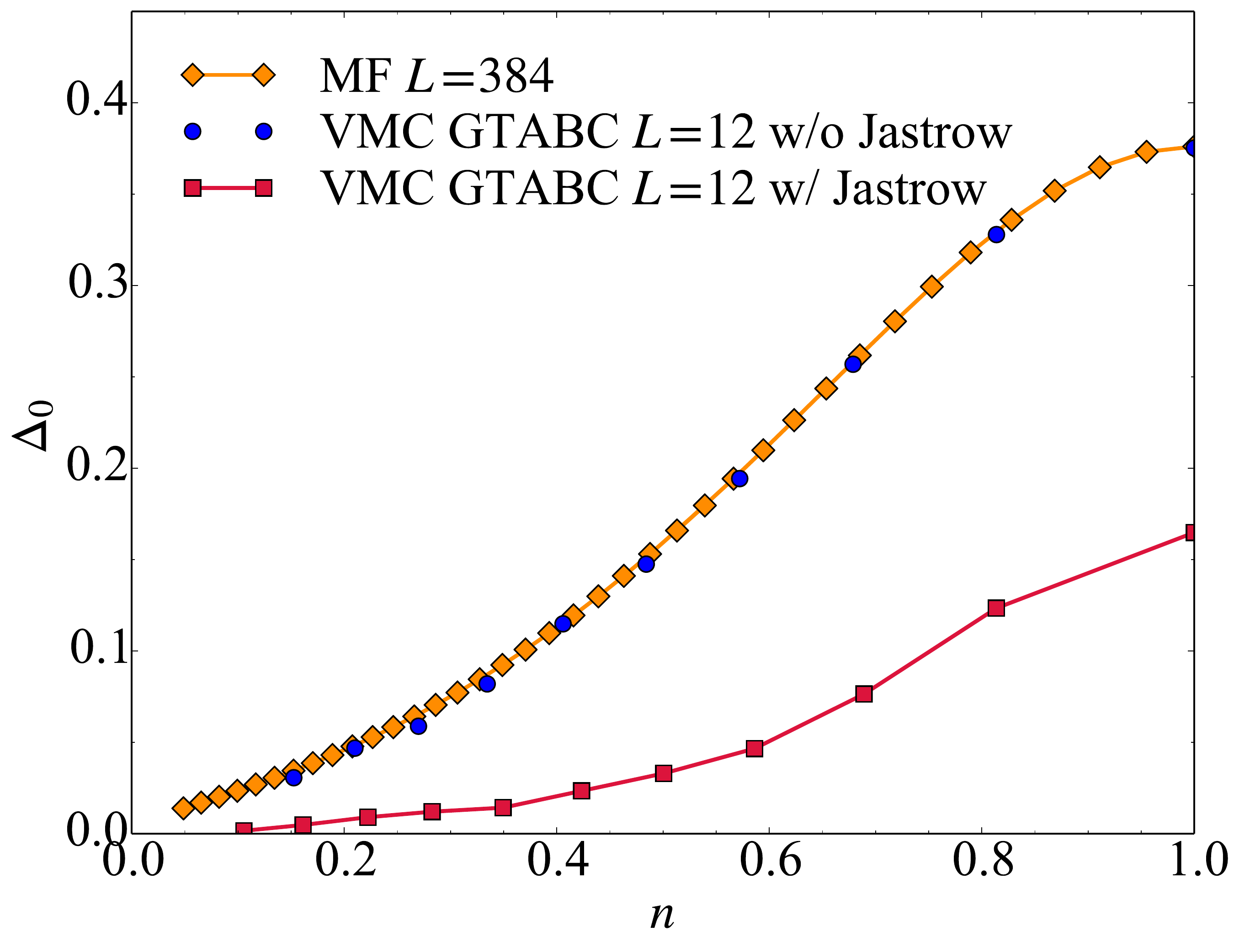}
  \caption{\label{fig:jastrow-mf} 
    $s$-wave variational parameter $\Delta_0$ as a function of $n$ at $U=-2$.  
    The results for  mean-field calculations on $L=384$ and 
    VMC on $L=12$ with and without Jastrow correlator in GCE with TABCs are shown. 
    The error bars in the VMC results are smaller than the symbol sizes.
  }
\end{figure}

Having confirmed the significant reduction of the finite-size effects,  
we show in Fig.~\ref{fig:jastrow-mf} how the Jastrow correlator affects the magnitude of the optimal variational parameter. 
By using the same system size of $L=12$ with the same number of twist angles, 
the Jastrow correlator reduces the magnitude of the $s$-wave variational parameter by more than a factor two for $n=1$. 
Even when the electron density $n$ is small, the reduction of the variational parameter is not negligible, 
showing the importance of the inclusion of the electron correlations. 
Note that, this systematic comparison of the variational wavefunction with and without Jastrow 
correlator for the entire doping range has been made possible only with GTABCs,  
because the results with a single boundary condition exhibit oscillatory behaviors both in 
the mean-field approximation and VMC.  

\subsection{pairing correlation function}

The finite order or variational parameters observed in the mean-field approximation or the VMC for finite-size 
systems are due to the wavefunction Ansatz that explicitly breaks the U(1) symmetry. 
In order to 
compare the VMC results with the ones of the numerically exact AFQMC, 
it is necessary  
to study  the off-diagonal long-range  order by 
computing superconducting correlation functions. To this purpose  
we consider the $s$-wave pairing correlation function 
\begin{equation}\label{eq:phi2}
  \phi^2(L) = 
  \frac{1}{2N}\sum_{i}
  \left \langle 
  \Delta_i^\dag \Delta_{i+j}
  + {\rm H.c.}
  \right \rangle,
\end{equation}
where $\Delta_i^\dag =   c_{i \uparrow}^{\dag} c_{i \downarrow}^{\dag}$ and, $i$ and  
$i+j$ are  sites  at  the maximum distance allowed by the boundary conditions 
of the cluster.

\begin{figure}[t!]
  \includegraphics[width=0.4\textwidth]{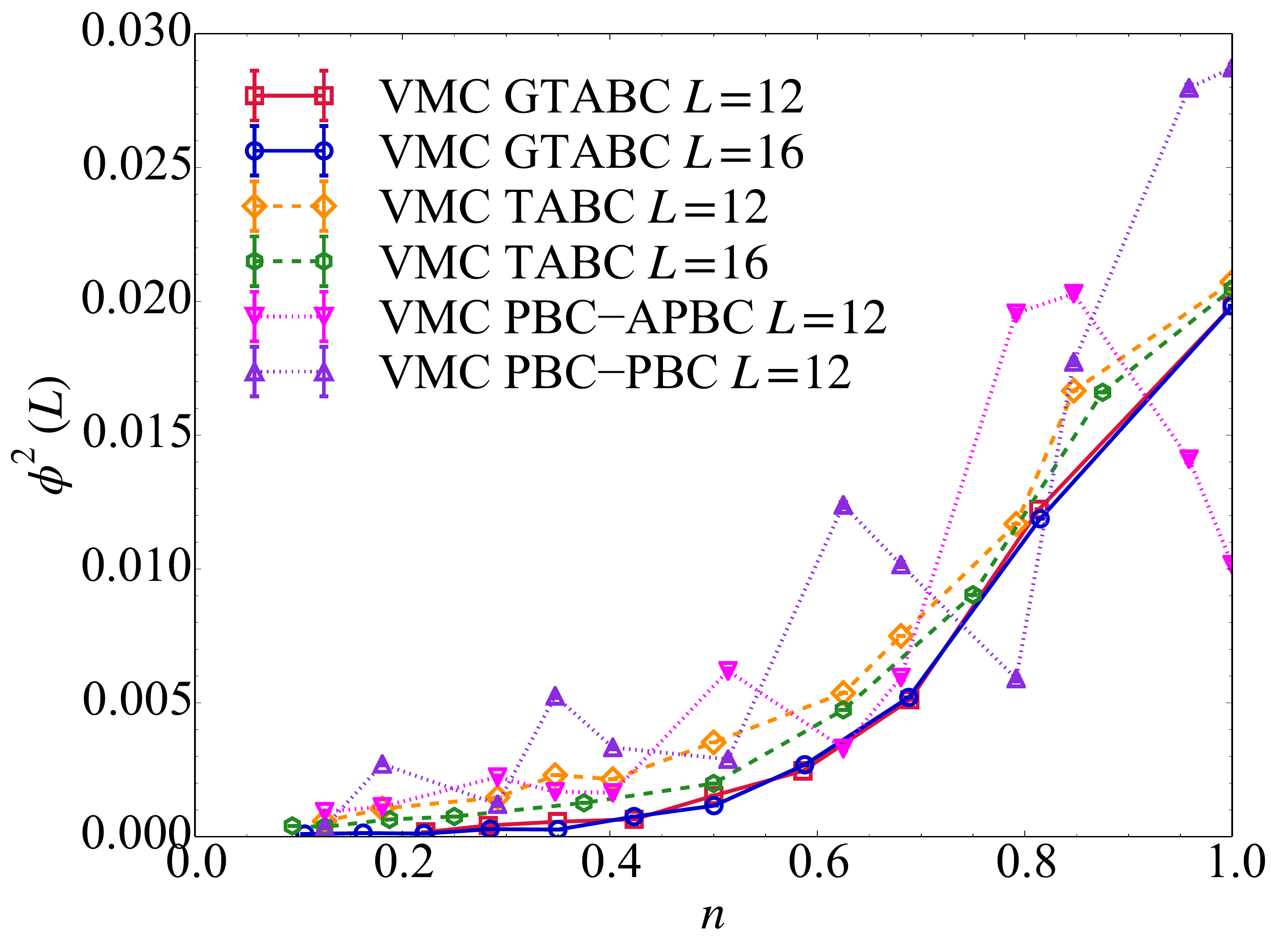}
  \caption{\label{fig:phi2U-2} 
    Pairing correlations $\phi^2$ as a function of $n$ at $U=-2$ calculated by VMC. 
    The system size and boundary conditions used are indicated in the figure. 
    The notations are the same as those in Fig.~\ref{fig:jastrow}. 
  }
\end{figure}

Figure~\ref{fig:phi2U-2} shows the calculated pairing correlation functions with VMC for various 
boundary conditions. As in the case of the variational parameter discussed in the previous section, 
strong finite-size effects are observed for $L=12$ with a single boundary condition. 
By increasing the system size to $L=16$, TABCs with CE reduce 
the oscillatory dependence as a function of $n$, but only with GCE the size effects become 
almost negligible within the available cluster sizes.

\begin{figure*}[t!]
  \includegraphics[width=1.0\textwidth]{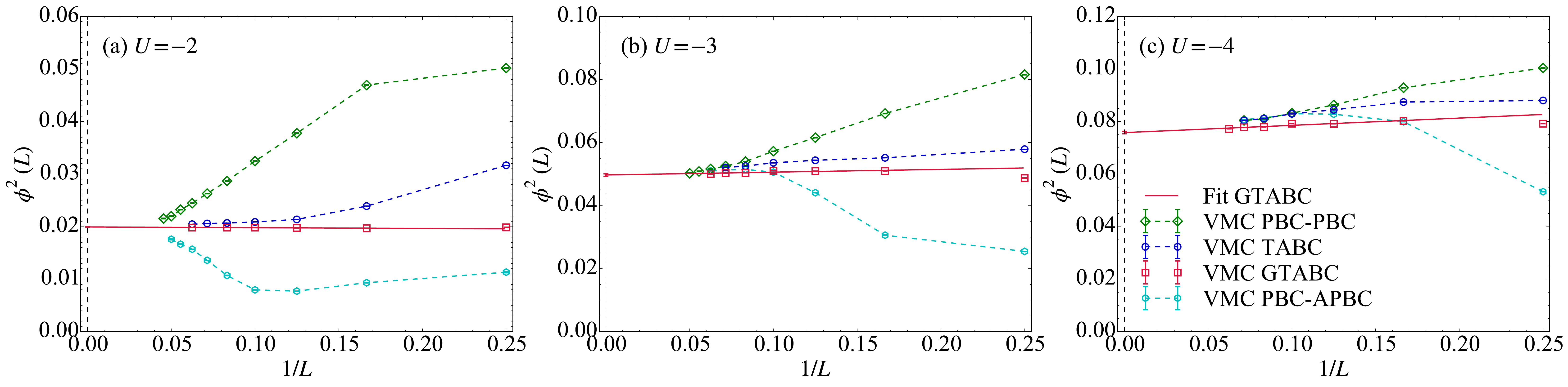}
  \caption{\label{fig:hfVMC} 
    Finite-size-scaling analyses of the pairing correlation $\phi^2(L)$ 
    for (a) $U=-2$, (b) $U=-3$, and (c) $U=-4$ by VMC 
    at half filling with different boundary conditions. 
    The solid lines are  fits to the GTABC data.
    The extrapolated values to the thermodynamic limit $\lim_{L\to \infty} \phi^2(L)$ 
    are indicated at $1/L=0$ for each panel and correspond to 
    \phihfUtwovmc for $U=-2$, 
    \phihfUthreevmc for $U=-3$, and 
    \phihfUfourvmc for $U=-4$.  
    The notations are the same as in Fig.~\ref{fig:jastrow}. 
  }
\end{figure*}

\begin{figure*}[t!]
  \includegraphics[width=1.0\textwidth]{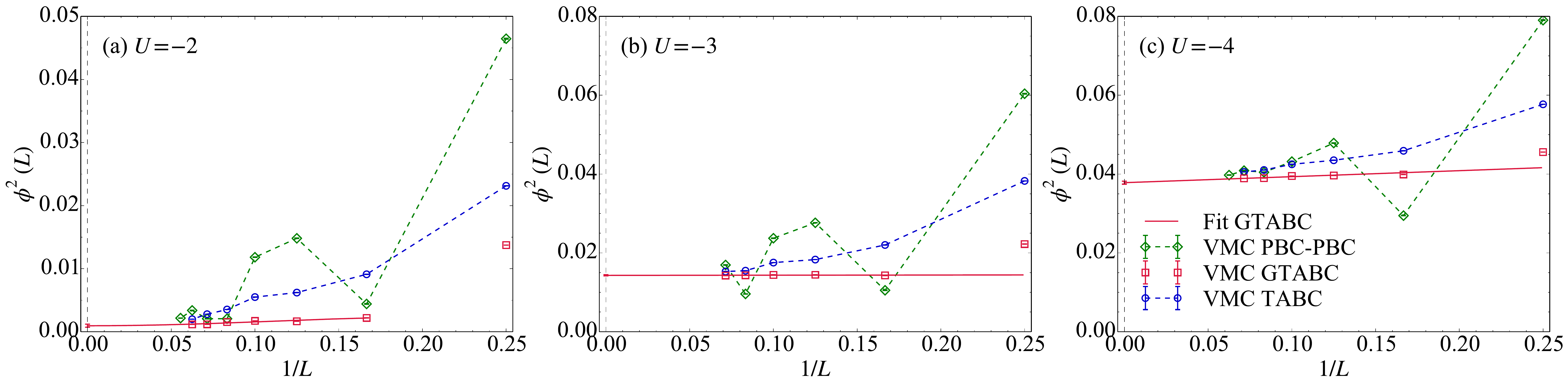}
  \caption{\label{fig:qfVMC}
    Finite-size-scaling analyses of the pairing correlation $\phi^2(L)$ 
    for (a) $U=-2$, (b) $U=-3$, and (c) $U=-4$ by VMC
    at quarter filling with different boundary conditions. 
    The solid lines are fits to the GTABC data. 
    The extrapolated values to the thermodynamic limit $\lim_{L\to \infty} \phi^2(L)$ 
    are indicated at $1/L=0$ for each panel and correspond to 
    \phiqfUtwovmc for $U=-2$, 
    \phiqfUthreevmc for $U=-3$, and 
    \phiqfUfourvmc for $U=-4$.   
    The notations are the same as those in Fig.~\ref{fig:jastrow}. 
  }
\end{figure*}

Careful finite-size-scaling analyses for the pairing correlation functions 
are done for $U=-2,-3,$ and $-4$ at half filling and at quarter filling 
in Figs.~\ref{fig:hfVMC} and ~\ref{fig:qfVMC}, respectively. 
We observe that, even at half filling, it is almost impossible to extrapolate the pairing correlations with a single twist since this approximation changes  behavior as the system size increases, especially when the value of the $|U|$ is small. Instead,  it is clearly evident  that the TABCs with GCE represents the best 
method to deal with finite size effects, also much better than TABCs within CE.
In particular, at quarter filling, 
severe system-size dependencies of the correlation functions are observed, 
implying that the finite-size scaling with a single twist is 
almost impossible. 

\begin{figure}[t!]
  \includegraphics[width=0.4\textwidth]{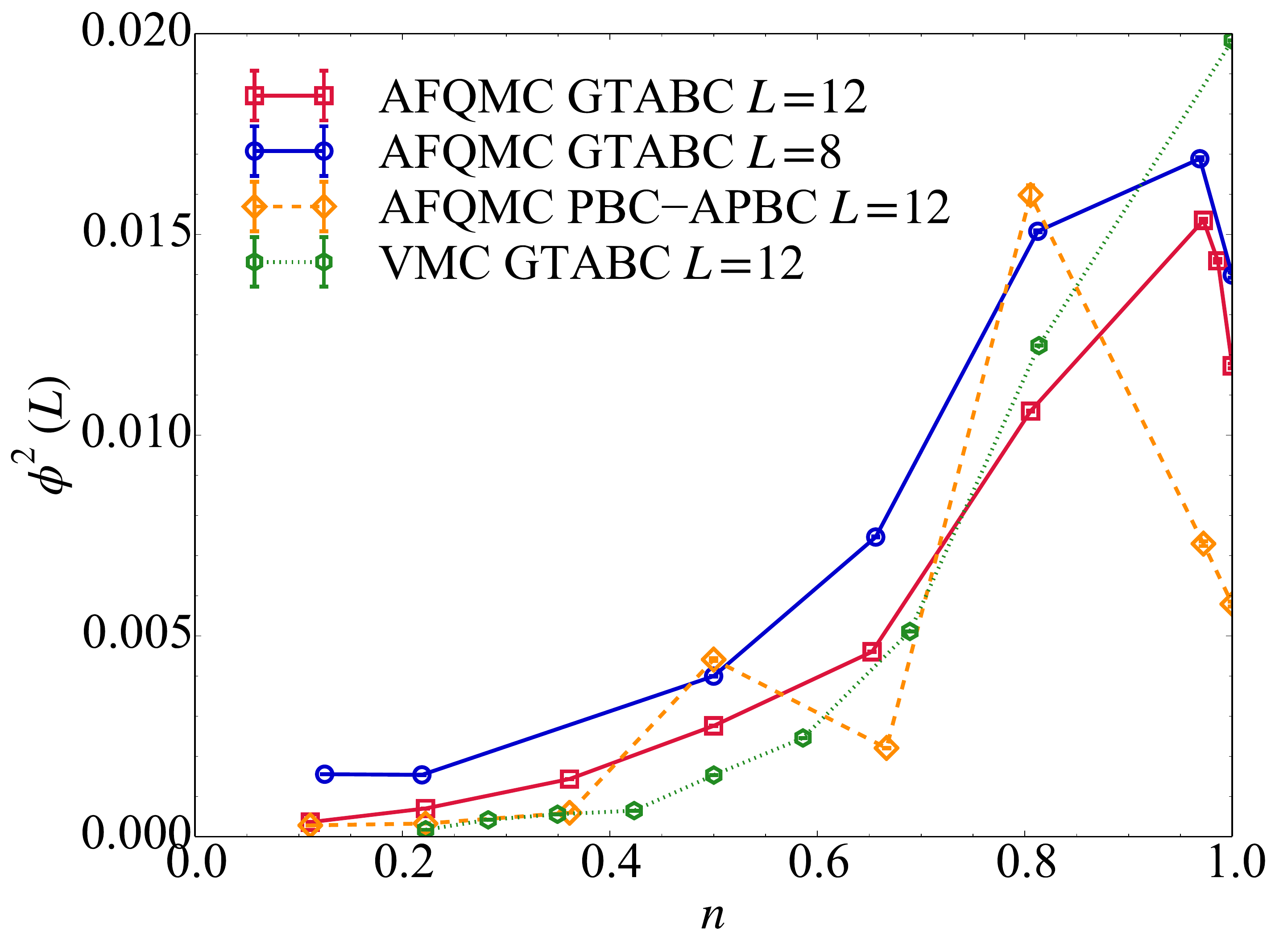}
  \caption{\label{fig:compafVMC}
    Pairing correlation $\phi^2$ as a function of $n$ at $U=-2$ 
    by AFQMC and VMC. The system size and boundary conditions are indicated in the figure.
    The notations are the same as in Fig.~\ref{fig:jastrow}.}
\end{figure}

Figure~\ref{fig:compafVMC} shows the pairing correlations obtained 
with AFQMC for $L=8$ and $L=12$ as well as VMC in GCE on $L=12$. 
As in the case of VMC, the PBC  results show significant size effects, 
that are reduced significantly by GTABCs also for AFQMC. 
Close to half filling, within AFQMC,  
the value of $\phi^2$ becomes larger than the corresponding one at half filling, a behavior qualitatively different from the one observed within VMC.   
This effect has been reported in the early QMC study of the negative-$U$ Hubbard model~\cite{Scalettar1989}, 
and can be attributed to the spin-flop transition in the strong-coupling limit, 
where the model, at small doping, can be mapped to the Heisenberg model  in presence of a small magnetic field (see also Sec.~\ref{Sec3D}). In this case as soon a non zero magnetic field is present the
order parameter ``flops'' in the xy-plane.

\begin{figure*}[t!]
  \includegraphics[width=1.0\textwidth]{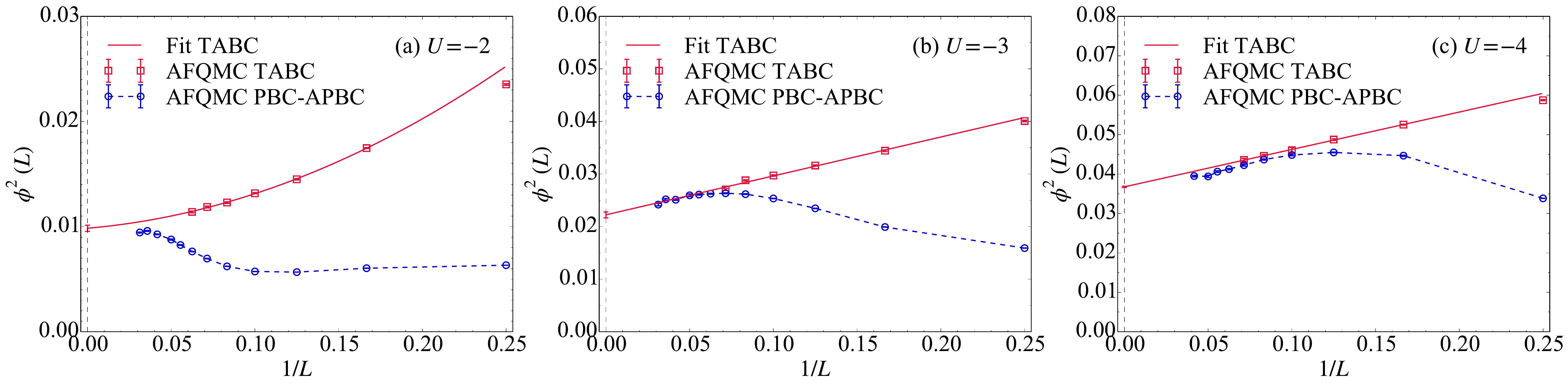}
  \caption{\label{fig:afHF} 
    Finite-size-scaling analyses of the pairing correlation $\phi^2(L)$ 
    for (a) $U=-2$, (b) $U=-3$, and (c) $U=-4$ by AFQMC 
    at half filling with different boundary conditions. 
    The solid lines are the fit to the TABC data. 
    The TABC extrapolated values to the thermodynamic limit $\lim_{L\to \infty} \phi^2(L)$ 
    are indicated at $1/L=0$ for each panel and correspond to 
\phihfUtwo for $U=-2$, 
\phihfUthree for $U=-3$, and 
\phihfUfour for $U=-4$. 
    The notations are the same as those in Fig.~\ref{fig:jastrow}. 
  }
\end{figure*}

\begin{figure*}[t!]
  \includegraphics[width=1.0\textwidth]{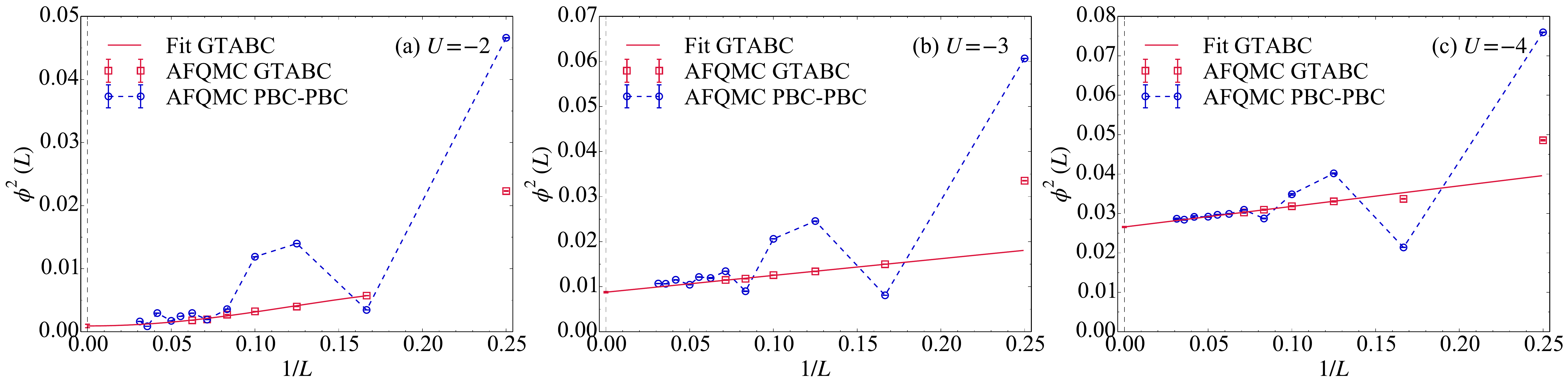}
  \caption{\label{fig:afQF} 
    Finite-size-scaling analyses of the pairing correlation $\phi^2(L)$ 
    for (a) $U=-2$, (b) $U=-3$, and (c) $U=-4$ by AFQMC 
    at quarter filling with different boundary conditions. 
    The solid lines are the fit to the GTABC data. 
    The GTABC extrapolated values to the thermodynamic limit $\lim_{L\to \infty} \phi^2(L)$ 
    are indicated at $1/L=0$ for each panel and correspond to
\phiqfUtwo for $U=-2$, 
\phiqfUthree for $U=-3$, and 
\phiqfUfour for $U=-4$.  
    The notations are the same as those in Fig.~\ref{fig:jastrow}. 
  }
\end{figure*}

Despite GTABCs, visible size effects remain in Fig.~\ref{fig:compafVMC} for the AFQMC case, and we have therefore focused on a few filling values, that we have systematically studied as a function of the system size.
We show the finite-size scaling of the pairing correlations at half filling and at quarter filling 
calculated by AFQMC 
in Figs.~\ref{fig:afHF} and \ref{fig:afQF}, respectively. 
As expected from the previous VMC study, also in the case of AFQMC calculations,
the TABCs with GCE allows a finite size scaling much 
better than the one with a single boundary condition. 
At half filling, the values of $\phi^2$ extrapolated to the thermodynamic limit are therefore computed with high accuracy, as shown in Fig.~\ref{fig:afHF}.

At quarter filling, the situation is even worse for the single twist 
approach, and 
severe system-size dependencies of the correlation functions prevent 
a systematic extrapolation to the thermodynamic limit. 
Fortunately this remains possible within TABC approach and  
controlled  extrapolations can be done also for AFQMC calculations. The thermodynamic values of superconducting correlations are therefore computed with high 
accuracy, as shown in Fig.~\ref{fig:afQF}.

\subsection{Attractive-repulsive mapping and order parameters}\label{Sec3D}
The negative-$U$ Hubbard model can be mapped to the positive-$U$ Hubbard model 
with the particle-hole transformation~\cite{Shiba1972}
\begin{eqnarray} \label{eq:pht}
  &\tilde{c}_{i\uparrow}   &:= c_{i \uparrow},  \\
  &\tilde{c}_{i\downarrow} &:= (-1)^{x_i + y_i} c_{i \downarrow}^\dag.
\end{eqnarray}
This mapping allows us to compare the results of the pairing correlation function $\phi^2(L)$ 
with those of the transverse spin-spin correlation function in the positive-$U$ Hubbard model.  
Indeed, in terms of the newly defined operators $\tilde{c}_{i\sigma}$, $\tilde{c}_{i\sigma}^\dag$ and 
$\tilde{n}_{i\sigma} = \tilde{c}_{i\sigma}^\dag \tilde{c}_{i\sigma}$, 
the Hamiltonian changes, up to a constant, to: 
\begin{eqnarray}
   H &=& -t \sum_{\langle i,j \rangle , \sigma} 
  \left(\tilde{c}^\dag_{i\sigma} \tilde{c}^\dagga_{j\sigma}   + {\rm H. c.} \right)   
  + |U| \sum_{i} \tilde{n}_{i\uparrow} \tilde{n}_{i\downarrow} \notag \\
  &-& \sum_{i} \left[ (\mu - U) \tilde{n}_{i\uparrow} - \mu \tilde{n}_{i\downarrow} \right],  
\end{eqnarray}
whereas $\phi^2(L)$ can be written as the transverse spin-spin correlation function 
\begin{eqnarray}
  M^2_{xy}(L)
  &=& \frac{1}{2N} \sum_{i} (-1)^{x_j + y_j} \left \langle S_{i}^+ S_{i+j}^- + {\rm H.c.} \right \rangle \notag \\
  &=& \frac{1}{N}  \sum_{i} (-1)^{x_j + y_j} \left \langle S_{i}^x S_{i+j}^x + S_{i}^y S_{i+j}^y \right \rangle, 
\end{eqnarray}
where 
$S_{i}^+ = \tilde{c}_{i\uparrow}^\dag \tilde{c}_{i\downarrow}$, $S_{i}^- = {(S_{i}^+)}^\dag$, 
$S_{i}^x = (S_{i}^+ + S_{i}^-)/2$, and 
$S_{i}^y = (S_{i}^+ - S_{i}^-)/2{\rm i}$. 
Similarly, the charge-charge correlations in the negative-$U$ Hubbard model 
can be mapped to the longitudinal spin-spin correlations in the positive-$U$ Hubbard model. 
In the present study, however, the charge-charge correlations 
are not considered as they will not dominate 
over the pairing correlations for large distances away from the half filling.

Since the negative-$U$ Hubbard model with $\mu = U/2$ (the half-filled case) corresponds to 
the positive-$U$ Hubbard model with zero magnetic field, 
the SU(2) symmetric staggered magnetization $M_0$ in the thermodynamic limit can be 
estimated from $M_{xy}(L)$ through the relation 
\begin{equation}
  M_0 
  = \sqrt{\frac{3}{2} \lim_{L\to\infty}  M^2_{xy}(L)},  
  \label{eq:M0}
\end{equation}
where the factor $3/2$ within the square root is included to take into account the contribution from 
the longitudinal spin-spin correlation which is not present in $M_{xy}^2(L)$. 
The estimated values of $M_0$ 
are reported in Table~\ref{table1}.
For $|U|=2$ and $4$, these values are in agreement 
with a recent study~\cite{SZhang2016} and, for $|U|=2$, also with a 
previous 
work~\cite{Sorella2015} by one of us. 
On the other hand an earlier work~\cite{PRX_Hub} reported a too small value of 
$M_0$ for $|U|=2t$, that was however affected by 
an error in the extrapolation to the thermodynamic limit~\cite{SZhang2016}. 

At quarter filling, the CDW order 
disappears and we are left to study only the $s$-wave order parameter defined 
as 
\begin{equation}
  \Phi_{s} = \sqrt{\lim_{L\to\infty}  \phi^2(L)}. 
  \label{eq:Ds}
\end{equation}
The estimated values of $\Phi_{s}$ 
from the extrapolated values of $\phi^2(L)$ are reported in Table~\ref{table2}. 

In order to test the accuracy of the variational wave function  in the thermodynamic limit, we have also compared the VMC estimates of the ground state energies with the  AFQMC ones in Table~\ref{table3} and Table~\ref{table4} at half filling and at quarter filling, respectively. The energies 
obtained via AFQMC in the present study are in agreement with the exact energies of previous works. It is worth mentioning  that VMC energies are providing quite good upper bounds to the exact energies, especially in the weak-coupling regime.

\begin{table}
  \caption{
    \label{table1} 
    Comparison of the $s$-wave superconducting (antiferromagnetic) order parameter 
    $M_0$ defined in Eq.~(\ref{eq:M0}) 
    for the negative (positive)-$U$ Hubbard model at half filling ($n=1$). 
    For the VMC case we have not included the factor $\sqrt{3 \over 2}$, see discussion in the conclusions.
    The number in each parenthesis in this work indicates 
    the uncertainty due to the extrapolation to the thermodynamic limit. 
    The AFQMC simulations in Ref.~\cite{Sorella2015} are performed with the modified-boundary conditions, 
    while in Refs~\cite{SZhang2016} and \cite{PRX_Hub} with the TABCs method. 
    DMET stands for density-matrix-embedding theory, and MF for the standard BCS mean-field theory.     
  }
  \begin{tabular}{lccc} 
    \hline
    \hline
    \multicolumn{4}{c}{$n=1$} \\
    \hline
    $|U|/t$ & 2  & 3 & 4 \\
    \hline 
    MF                       & \HFtwo    & \HFthree   & \HFfour \\
    VMC   (this work)        & \MUtwovmc & \MUthreevmc & \MUfourvmc \\
    AFQMC (this work)        & \MUtwo    & \MUthree & \MUfour \\
    AFQMC \cite{Sorella2015} & 0.120(5)  & --       & --       \\
    AFQMC \cite{SZhang2016}  & 0.119(4)  & --       & 0.236(1) \\
    AFQMC \cite{PRX_Hub}     & 0.094(4)  & --       & 0.236(1) \\
    DMET  \cite{PRX_Hub}     & 0.133(5)  & --       & 0.252(9) \\
    \hline
    \hline
  \end{tabular}
\end{table}

\begin{table}
  \caption{
    \label{table2} 
    Comparison of the $s$-wave superconducting order parameter $\Phi_s$ defined in Eq.~(\ref{eq:Ds}) 
    for the negative-$U$ Hubbard model at quarter filling ($n=0.5$). 
    The number in each parenthesis in this work indicates 
    the uncertainty due to the extrapolation to the thermodynamic limit. 
  }
  \begin{tabular}{lccc} 
    \hline
    \hline
    \multicolumn{4}{ c }{$n=0.5$} \\
    \hline
    $|U|/t$ & 2  & 3 & 4 \\
    \hline 
    VMC     (this work) & \MxyUtwovmc & \MxyUthreevmc & \MxyUfourvmc \\
    AFQMC   (this work) & \MxyUtwo & \MxyUthree & \MxyUfour \\
    \hline
    \hline
  \end{tabular}
\end{table}

\begin{table}
  \caption{
    \label{table3} 
    Comparison of the ground state energies
    for the negative (positive)-$U$ Hubbard model at half filling ($n=1$). 
    The number in each parenthesis in this work indicates 
    the uncertainty due to the extrapolation to the thermodynamic limit. 
  }
  \begin{tabular}{lccc} 
    \hline
    \hline
    \multicolumn{4}{c}{$n=1$} \\
    \hline
    $|U|/t$ & 2  & 3 & 4 \\
    \hline 
    VMC   (this work)        & \Enhftwovmc & \Enhfthreevmc & \Enhffourvmc \\
    AFQMC (this work)        & \Enhftwo    & \Enhfthree & \Enhffour \\
    AFQMC \cite{Sorella2015} & -2.175469(92)  & -2.501412(52)   &  --      \\
    AFQMC \cite{SZhang2016}  & -2.1760(2)  & --       & -2.8603(2) \\
    AFQMC \cite{PRX_Hub}     & -2.1763(2)  & --       & -2.8603(2) \\
    DMET  \cite{PRX_Hub}     & -2.1764(3)  & --       & -2.8604(3) \\
    \hline
    \hline
  \end{tabular}
\end{table}

\begin{table}
  \caption{
    \label{table4} 
    Comparison of the ground state energies  
    for the negative-$U$ Hubbard model at quarter filling ($n=0.5$). 
    The number in each parenthesis in this work indicates 
    the uncertainty due to the extrapolation to the thermodynamic limit. 
  }
  \begin{tabular}{lccc} 
    \hline
    \hline
    \multicolumn{4}{ c }{$n=0.5$} \\
    \hline
    $|U|/t$ & 2  & 3 & 4 \\
    \hline 
    VMC     (this work) & \Enqftwovmc & \Enqfthreevmc & \Enqffourvmc \\
    AFQMC   (this work) & \Enqftwo & \Enqfthree & \Enqffour \\
    \hline
    \hline
  \end{tabular}
\end{table}

\section{Conclusions and discussions} \label{sec4}
To conclude, 
finite-size effects on the $s$-wave order parameter and 
pairing correlation functions have been studied in details for the negative-$U$ Hubbard model with VMC and AFQMC methods.  
GTABCs reduce systematically 
the finite size effects and   provide smooth extrapolation to the thermodynamic limit. 
This has enabled us to obtain well converged  results on energy and order parameter for several values of 
$U/t$ and doping, and to study very efficiently  
the physical properties in the thermodynamic limits 
of  our  Jastrow correlated wavefunction  as a function of doping. 
Indeed, we have shown that the Jastrow correlator in VMC significantly reduces the magnitude of the $s$-wave variational parameter in the entire doping range, already at $U=-2$. 

We have also presented the comparison of the pairing correlation functions 
obtained by VMC and by the numerically exact AFQMC. 
In this case  VMC is in  good agreement with AFQMC for finite doping.
At half filling, there exists  a pseudo-SU(2) symmetry
defined by the SU(2) rotations in spin space applied to the Hamiltonian after the particle-hole 
transformation in Eq.~(\ref{eq:pht}), that remains therefore  invariant and commuting 
with the pseudo spin operators [the spin operators after the particle-hole transformation of Eq.~(\ref{eq:pht})].
This symmetry is clearly
accidental, as  
is no longer satisfied  
by the inclusion of a tiny next-nearest-neighbor hopping $t'$~\cite{Otsuka2018}.
Since our variational  wavefunction 
breaks this accidental pseudo-SU(2) symmetry  
the agreement between the VMC and the necessarily symmetrical (as the exact  ground state in any finite lattice is  
a singlet after particle-hole transformation~\cite{lieb}) AFQMC  
in this case is not very good 
 just because in the VMC the order is only in one  of the possible 
directions  of a three-component order parameter.  
For this reason, when comparing VMC and AFQMC in Table~\ref{table1}, 
we have not included in the VMC entries the factor $\sqrt{3/2}$ implied by the definition of $M_0$ in Eq.~(\ref{eq:M0}), as we have verified that, within VMC, the order is  in the $xy$ plane,
because the  CDW order, corresponding, in the positive-$U$ language,  to the $z$ component of the antiferromagnetic order-parameter, is always negligible.
This consideration  
explains also why the spin-flop transition observed in AFQMC --  $M_{xy}$ jumps to a larger 
value as soon as we depart from half-filling --  
is not present in VMC, because, as shown in Fig.~\ref{fig:compafVMC},  $M_{xy}$ 
appears a smooth and  monotonically decreasing function of the doping.

Apart for symmetry considerations, that can be restored by standard symmetry projection techniques\cite{Imada,Scuseria}, our wavefunction can be also improved, 
for example, by taking into account the backflow correlations~\cite{Holzmann2003}, 
as it was done in the positive-$U$ Hubbard model~\cite{Tocchio2008,Tocchio2011}.

The method for reducing finite-size effects, 
developed in this paper, is applicable for any correlated lattice model. 
The calculation in GCE will be particularly useful for 
investigating the doping dependence of the $d$-wave superconductivity 
in the positive-$U$ Hubbard model with parameters relevant for cuprates. 
Furthermore, the reduction of the order parameter in the entire doping range 
due to the Jastrow factor suggests that the electron-correlation  
are  not be negligible even for 
weakly attracting fermions in the low-electron-density regime.  
This implies that the method will be also promising for  studying 
the ground-state properties of dilute electron systems. 
Such systems may include TiSe$_2$ in the series of transition metal dichalcogenides~\cite{Monney2010,Rossnagel2011,Watanabe2015,Kaneko2018}, 
where its electronic state is in vicinity of the semimetal-semiconductor transition  
and considered to be a candidate of excitonic insulators, 
in which coherent electron-hole pairs are formed and condensate spontaneously~\cite{Jerome1967,Halperin1968}.

\acknowledgements
We ackowledge Federico Becca for useful discussions.
Computational resources were provided mostly by CINECA-Pra133322.
A part of computations has been done on HOKUSAI GreatWave supercomputer 
at RIKEN Advanced Center for Computing and Communication (ACCC). 
K. S. acknowledges support from the JSPS Overseas Research Fellowships.


%

\end{document}